\documentclass[12pt]{article}
\usepackage{graphicx}
\usepackage[hang,small,bf]{caption}
\newcommand{\be}[3]{\begin{equation}  \label{#1#2#3}}
\newcommand{\bea}[3]{\begin{eqnarray}  \label{#1#2#3}}
\newcommand{\ee}{\end{equation}}
\newcommand{\eea}{\end{eqnarray}}
\newcommand{\ba}{\begin{array}}
\newcommand{\ea}{\end{array}}
\renewcommand{\arraystretch}{1.8}

\let\LARGE=\Large
\let\Large=\large
\let\large=\normalsize

\usepackage{amssymb,bbold,euscript}

\begin{document}


\thispagestyle{empty}

\begin{flushright}
\hfill{AEI-2003-098}\\
\hfill{UUPHY/2003-11} \\
\hfill{hep-th/0312063}
\end{flushright}

\vspace{15pt}

\begin{center}{ \LARGE{\bf
De Sitter vacua from $N=2$ gauged supergravity }}

\vspace{30pt}

{\bf Klaus Behrndt}$^a$
and {\bf Swapna Mahapatra}$^b$

\vspace{20pt}

$^a$ {\it Max-Planck-Institut f\"ur Gravitationsphysik,
Albert Einstein Institut \\
Am M\"ulenberg 1, 14476 Golm,
Germany}\\[1mm]
{e-mail: behrndt@aei.mpg.de}

\vspace{10pt}

$^b$ {\it Physics Department, Utkal University, 
Bhubaneswar 751 004, India} \\[1mm]
{e-mail: swapna@iopb.res.in}

\vspace{50pt}

{ABSTRACT}

\end{center}
Typical de Sitter (dS) vacua of gauged supergravity correspond to
saddle points of the potential and often the unstable mode runs into a
singularity. We explore the possibility to obtain dS points where the
unstable mode goes on both sides into a supersymmetric smooth
vacuum. Within $N=2$ gauged supergravity coupled to the universal
hypermultiplet, we have found a potential which has two supersymmetric
minima (one of them can be flat) and these are connected by a de Sitter
saddle point. In order to obtain this potential by an Abelian
gauging, it was important to include the recently proposed quantum
corrections to the universal hypermultiplet sector. Our results apply
to four as well as five dimensional gauged supergravity theories.

\newpage


\section{Introduction}

The conjectured AdS/CFT duality \cite{900} has led to renewed intense
research in gauged supergravity theories in various dimensions.  The
study of $N=2$ gauged supergravity theories has particularly enriched
our understanding regarding many interesting aspects of domain wall
solutions and the associated renormalization group (RG) flow in a dual
superconformal field theory \cite{801, 905}.  These domain wall
solutions have also played an important role in the context of brane
world scenario proposed as an alternative to compactification
\cite{906}.

Type-II or M-theory when compactified on a Calabi-Yau threefold gives
rise to $N=2$ models with vector and hypermultiplets in four and five
dimensions respectively. The scalars belonging to the vector and
hypermultiplets are known to parametrize a manifold that is a product
of a special K\"ahler and a quaternionic manifold.  The various
gauging procedures involving gauging of the Abelian $U(1)_R$ symmetry,
the full $SU(2)_R$ symmetry, isometries of the vector and the
hypermultiplets \cite{907, 911, 913, 915, 916, 917} provide an
explicit derivation of the superpotential and the associated scalar
potential. From the analysis of the critical points of the potential,
one has several instances where the potential admits supersymmetric
extrema. Explicit domain wall solutions have been obtained which
interpolate between different extrema of the potential.  In order to
make contact with string or M-theory, one is especially interested in
potentials that are related to flux or Scherk-Schwarz reductions (or
massive $T$-dualities) \cite{260}.

Besides supersymmetric extrema, the potentials can also exhibit de
Sitter extrema so that supersymmetry is spontaneously broken in these
vacua. The exploration of de Sitter vacua in string or M-theory has
attracted much attention in recent times, not only from the point of
view of gauged supergravity (or hyperbolic reductions) \cite{180, 160,
210, 150,100, 211}, but also specific brane configurations have been
investigated from various perspectives \cite{240,120,110}. The
potential obtained by explicit supersymmetry breaking (e.g.\ by
addition of anti branes) have more room for phenomenologically
interesting quantities as the strongly constrained potentials obtained
in gauged supergravity. For example, de Sitter vacua in gauged sugra
typically correspond to saddle points (with some notable exceptions as
e.g.\ in \cite{100}) and often do not fulfil the slow roll requirement
\cite{150, 333}, which ensures sufficient inflation.  Therefore, these
potentials can only contribute to fast roll inflation as proposed in
\cite{190}, which might nevertheless occur during some period of the
cosmological evolution. On the other hand, a saddle point can give a
so-called locked inflation, where the scalar field oscillates along the
stable direction with an amplitude much larger than the curvature
scale of the unstable direction \cite{931}, see also \cite{140}.

An especially rich vacuum structure in gauged supergravity is expected
to arise if one gauges the isometries of the quantum corrected moduli
space.  Only very few quantum corrections have been explicitly
calculated. In some cases the correct geometry of these spaces follows
by imposing lower dimensional supersymmetry combined with certain
number of isometries respected by the quantum corrections \cite{918,
919, 921, 922}.  For example in the double-tensor calculus \cite{923},
the gauge symmetries ensure that the 4-dimensional quaternionic space
of the universal hypermultiplet (UH) has at least two commuting
isometries. The explicit construction of these spaces is a fairly
challenging task, but fortunately, the most general four dimensional
quaternionic metric with a $U(1) \times U(1)$ isometry group has been
found by Calderbank and Pedersen \cite{924}, which
encodes all perturbative and non-perturbative corrections that respect
this symmetry.  Besides a non-perturbative instanton sum, this metric
includes the 1-loop correction to the universal hypermultiplet moduli
space, which has been recently verified by explicit calculation in the
nice paper \cite{921}, that moreover shows that all higher loop
corrections correspond to field redefinitions and that the instanton
sum has to be related to D2/M2-brane instanton, see also
\cite{922}. Additional D4 or 5-brane instantons would break all
isometries and may leave only discrete symmetries and hence are not
described by this metric.

Therefore, the investigation of this moduli space from the point of
view of gauged supergravity, opens the window of understanding of
non-perturbative physics in the vacuum structure. Supersymmetric 
vacua have been investigated already in \cite{915} and we want to 
focus in this paper on possible de Sitter vacua coming from this metric. 
To keep the expressions as simple as possible we consider only this
universal hypermultiplet and couple it the $N$=2 supergravity. 
In our analysis we do not consider any vector multiplets and therefore
our results apply to four as well as five dimensional $N$=2 supergravity, 
as in both cases the hypermultiplet moduli space is given by
a quaternionic space. But of course, from cosmological point of view,
one might prefer to use our results in the 4 dimensions and in most
applications we implicitly assume to be in 4 dimensions.

We shall gauge a linear combination of the two Abelian isometries and
find besides the supersymmetric extrema, a saddle point of the
potential that corresponds to a de Sitter vacuum. The unstable mode flows
on both sides to supersymmetric vacua and for a proper choice of
parameter one of them is flat and the other one is anti de Sitter.
Since there is no supersymmetric flow between two flat space vacua,
the de Sitter saddle point has to have an AdS vacuum on one
side. Note, the absence of vector multiplets implies that the BPS
vacua should not break supersymmetry, because the breaking of $N$=2 to
$N$=1 supersymmetry requires a massive spin 3/2 multiplet which has to
contain {\em two} vectors and therefore requires at least one vector
multiplet, see e.g.\ in \cite{230}. But because this metric has two
commuting isometries it is straightforward to gauge one of them with
the graviphoton and the second one with a vector of an additional
vector multiplet.

The organization of the paper is as follows: in Section 2, we
summarize relevant aspects of $N=2$ gauged supergravity and
quaternionic spaces, in Section 3, we discuss the Calderbank-Pedersen
metric and its relation to quantum corrected universal hypermultiplet
moduli space.  Section 4 contains the details of the gauging, the
superpotential and the analysis of the critical points. In Section 5,
we discuss the cosmological flow from the de Sitter saddle point to
the flat space minimum and finally in section 6 we present a
short summary and outlook.


\section{$N=2$ gauged supergravity and quaternionic spaces}


The field content of the bosonic sector of $N=2$ supergravity coupled
to $n_H$ hypermultiplets are the graviton $e^a_{\mu}$, the
graviphoton $A_{\mu}$ and the $4n_H$ hyperscalars $q^m$. We will
primarily be interested in the 4-dimensional case, but our results
will also hold in 5 dimensions since we do not consider couplings to
vector multiplets. Moreover, if we ignore the graviphoton, which will
play no role here, the bosonic part of the Lagrangian is then given
by,
\be201
e^{-1}{\cal L}^{N=2}_{bosonic} = -\frac{1}{2}R -
 \frac{1}{2}g_{mn}\nabla_{\mu}q^m\nabla^{\mu} q^n - g^2 V 
\ee
where, the scalar potential $V$ reads as,
\be202
V = - 3 \, P^i P^i + 2 \, k^m k^n g_{mn}(q)
\ee
and the covariant derivative of the hypermultiplet scalars is
given by,
\be203
\nabla_{\mu}q^m = \partial_{\mu}q^m + g A_{\mu}k^m(q) \ .
\ee
In the above, $g_{mn}$ is the metric of the $4$ dimensional
quaternionic space, $k^m(q)$ are the Killing vectors of the gauged
isometries of the quaternionic space with the gauge coupling constant
$g$ and $P^i(q)$ ($i = 1,2,3$) are the corresponding triplet of
Killing prepotentials.

Quaternionic-K\"ahler spaces allow for three (almost) complex
structures $J^i$ ($i = 1,2,3)$ defined by the algebra,
\be204
J^i \cdot J^j  = - \delta^{ij} + \epsilon^{ijk} J^k \ .
\ee
Denoting the quaternionic vielbein by $e^m$, one obtains
the triplet of 2-forms $\Omega^i$ as,
\be205
\Omega^i  =  -\frac{\kappa}{2} e^m \wedge J^i_{mn} e^n
\ee
where, $m, n = 1,2,3,4$ are the quaternionic indices. The holonomy
group of a $4n$ dimensional quaternionic manifold is contained in
$Sp(1) \times Sp(n)$. For $n=1$ this statement can be replaced by the
requirement that the Weyl-tensor of a $4$-dimensional quaternionic
space has to be (anti)-self-dual ${\it i.e.}$
\be206
W + \star W  = 0 \ .
\ee
For a quaternionic space in any dimension, the triplet of
$2$-forms $\Omega^i$ is expressed in terms of the $SU(2)$-part of the
quaternionic connection $A^i$ (not to be confused with the graviphoton!)
\be207
d A^i + \frac{1}{2} \epsilon^{ijk} A^j\wedge A^k = \Omega^i
\ee
which ensures that the triplet of two forms are covariantly
constant with respect to the $SU(2)$ connection, {\it i.e.}
\be208
d\Omega^i + \epsilon^{ijk} A^j\wedge \Omega^k = 0
\ee
and can be obtained from the spin connection 1-form $w^{mn}$ of the
quaternionic space
\be218
A^i = \frac{1}{2} w^{mn} J^i_{mn}
\ee
In the same way, the self-dual part gives the $Sp(n)$ connection.

Any quaternionic space is Einstein and as hypermultiplet moduli space,
it has to have negative scalar curvature.  Our complex structures are
anti-self-dual ($J^i_{mn} = -\frac{1}{2}\epsilon_{mnpq} J^i_{pq}$) so
that the triplet of $2$-forms can be written as,
\renewcommand{\arraystretch}{1.4}
\bea209
\Omega^1 &=& (e^1\wedge e^4 - e^2 \wedge e^3)\ , \nonumber \\
\Omega^2 &=& (e^1\wedge e^3 + e^2\wedge e^4) \ , \nonumber \\
\Omega^3 &=& (-e^1\wedge e^2 + e^3 \wedge e^4) 
\eea
where $e^m$ are the vielbeine for the quaternionic metric $g_{mn}$.
The isometries of this quaternionic manifold are generated by the
Killing vectors $k^m_I(q)$
\be219
\delta q^m = \epsilon^I k^m_I(q) \ .
\ee
In component notation, the Killing vectors can be expressed in terms of
the $SU(2)$ triplet of Killing prepotentials $P^i_I(q)$ as
\be221
k^n_I \Omega^i_{mn} = \nabla_m P^i_I = \partial_m P^i_I +
\epsilon^{ijk} A^j P^k_I
\ee
where, $i=1,2,3$ are the $SU(2)$ indices, $m,n$ are the quaternionic
indices and $I$ labels different isometries. Using the above
relation, one can write the prepotentials as,
\be222
P^i_I =  \nabla_m (k_I)_n (\Omega^i)^{mn} \ .
\ee
With these prepotentials, we introduce the scalar superpotential
$W$ by,
\be223
W =  {\sqrt{P^i P^i}} 
\ee
such that the potential (in the real notation) can be expressed in the
form,
\be224
V = 3 \, \Big( \, {2 \over 3} g^{mn}\partial_m W \partial_n W -  W^2 \Big) \ .
\ee
%


\section{Universal hypermultiplet and Calderbank-Pedersen metric}


The universal hypermultiplet arising in every Calabi-Yau
compactification of type IIA/M-theory e.g., contains the dilaton
$\phi$, the axion $D$ and a complex Ramond-Ramond scalar field $C$.
The corresponding moduli space receives both a 1-loop as well as
non-perturbative quantum corrections.  In this section, we want to
consider the quantum corrected moduli space.

Classically, the universal hypermultiplet parametrizes the
quaternionic-K\"ahler space ${SU(2,1)\over U(2)}$, which is a
symmetric coset space and is parametrized by the complex NS-NS and R-R
fields $S$ and $C$ respectively or alternatively in terms of four
real scalars. In type-IIA theory, the real part of the NS-NS field $S$
is related to the four dimensional dilaton and the imaginary part is
related to the axion $D$ which is dual to the three form $H$.  Loop
corrections to the universal hypermultiplet has also been discussed
earlier \cite{918}. The recent calculation of ref.\cite{921} shows
that nontrivial perturbative quantum corrections appear only at the
1-loop order and render the space non-K\"ahlerian.  They have been
obtained by considering possible deformations of the quaternionic
target space metric which preserve the Heisenberg subgroup of the
$SU(2,1)$ isometry group of classical moduli space. This subgroup is
generated by the classical Peccei-Quinn shift symmetries, which
should not be broken at the perturbative level. Moreover, they 
have shown that all higher loop corrections can be absorbed into field
redefinitions.

This immediately raises the question about non-perturbative
corrections, which in general is a notoriously difficult question. 
But
one may start with a subclass of deformations which leave invariant
certain symmetries as e.g.\ the $U(1) \times U(1)$ subgroup, which
appears e.g.\ as gauge symmetry in the double tensor calculus
\cite{923}. Assuming this symmetry, Calderbank-Pedersen were able to
find the most general (quaternionic) metric \cite{924}, which
reproduces exactly the 1-loop corrections found by Antoniadis etal.\
\cite{921}. They moreover argue that further corrections represent
a summation over all D2/M2 instantons (in the type IIA/M-theory
setting); see also \cite{922} for related proposals. In addition,
if one also
includes D4- and 5-brane instantons, one cannot expect the
resulting moduli space to have any isometries. This however, would exclude
any gauging of isometries and therefore the hyperscalars do not enter
the potential and it is then unclear how one can fix these moduli 
(note however, due to back reaction the corresponding internal space 
may not have the moduli that need to be fixed).

The anti selfdual Einstein metric of Calderbank and Pedersen (CP) can
be written in the form \cite{924},
\renewcommand{\arraystretch}{1.4}
\bea131
{ds}_{CP}^2 & = & {F^2 - 4\rho^2(F_\rho^2 + F_\eta^2)\over 4F^2}
\Big ({d\rho^2 + d\eta^2 \over \rho^2} \Big ) + \nonumber \\
&& {\Big [(F - 2\rho F_\rho)\alpha - 2\rho F_\eta\beta \Big ]^2 +
\Big [ (F + 2\rho F_\rho)\beta - 2\rho F_\eta\alpha \Big ]^2
\over F^2\Big [ F^2 - 4\rho^2 (F_\rho^2 + F_\eta^2)\Big ]}
\eea
where, the 1-forms $\alpha$ and $\beta$ are
\be132
\alpha = {\sqrt\rho}d\phi \quad , \qquad ~~ \beta = {(d\psi + \eta d\phi)
\over {\sqrt\rho}} 
\ee
and the two commuting isometries are given by the Killing vectors:
$\partial_{\phi}$ and $\partial_\psi$.  In another form, the same
metric becomes
\renewcommand{\arraystretch}{0.8}
\be100
ds_{CP}^2 = {1 \over F^2(\rho, \eta)} \left[ \det Q \, 
{d \rho^2 + d \eta^2 \over 4 \rho^2}
+ {1 \over \det Q} (d\phi\, , \ d\psi) \, N^t Q^2 N \,\left(\ba{c} d\phi \\ 
d\psi \ea \right) \right]
\ee
with
\be110
Q = \left(\ba{cc} {1 \over 2} F - \rho \partial_\rho F & 
- \rho \partial_\eta F \\ - \rho \partial_\eta F & 
{1 \over 2} F + \rho \partial_\rho F \ea \right) 
\quad , \qquad N = {1 \over \sqrt{\rho}} 
\left( \ba{cc} \rho & 0 \\ \eta & 1 \ea \right) \ .
\ee
This metric has positive scalar curvature if $\det Q > 0$. For 
$ \det Q < 0$, \, ($-g_{CP}$) is an anti-selfdual Einstein metric with 
negative scalar curvature $R(-g_{CP}) = -12$.  The point where $\det Q =
0$ is a curvature singularity whereas a zero of $F$ is a conformal
infinity (end of a given coordinate space).  As we will see below, one
can find a choice of parameter so that the curvature singularity, that
separates the positively and negatively curved quaternionic space, is
not there in a given coordinate region. 

The metric is completely specified by the function $F(\rho, \eta)$,
which is a real function of two variables $\eta$ and $\rho$ and obeys
the equation
\be133
\rho^2(\partial_\rho^2 + \partial_\eta^2) \, F(\rho, \eta) = 
{3\over 4} \, F(\rho, \eta)
\ee
i.e.\ it is an eigenfunction of the two dimensional Laplace-Beltrami
operator for the hyperbolic metric ${d\rho^2 + d\eta^2 \over \rho^2}$
with eigenvalue ${3 \over 4}$. This linear equation is invariant under 
$SL(2,R)$ transformation of
\be521
\tau = \eta + i\, \rho \quad \rightarrow \quad {a \tau + b \over c\tau +d}
\quad ,\qquad ad - bc=1
\ee
and hence the most general solution is given by a modular function, as
discussed below. 

If we first ignore the $\eta$-dependence, the basic solutions to the
eigenvalue equation are given by power functions
\be136
F_s(\rho, \eta) = \rho^s , \, ~~~~ s = \, {3\over 2}\, , \ -{1\over 2} \ .
\\
\ee
With the field identifications
\be139
\rho^2 = e^{-2\phi}, ~~~ C = C_1+iC_2 = \eta + {i\over 2}\phi,
~~~ \psi = D - C_1 C_2 \\
\ee
one can check that $F(\rho, \eta) = \rho^{3/2}$ gives exactly the tree
level result, i.e.\ the CP metric reduces the known classical
universal hypermultiplet metric. As shown in \cite{921} the
1-loop correction to this moduli space can be reproduced by
adding both solutions:
\be340
F(\rho, \eta) = \rho^{3/2} - \hat\chi \, \rho^{-1/2} \ .
\ee
The explicit calculation yields: $\hat\chi = - {4 \zeta(2) \chi \over
(2 \pi)^3}$, where $\chi$ is the Euler number of the Calabi-Yau space
and due to this 1-loop correction, the metric becomes non-K\"ahler. 
Moreover it has a singularity at $\rho^2 = \hat \chi$ where $\det Q =
0$. Note, the field redefinition (\ref{139}) receives also corrections
\cite{921} so that the singularity appears at strong coupling
$e^{-2\phi_4} \rightarrow \infty$ .  At this singularity we cannot
trust anymore the approximation and have to include further
corrections, i.e.\ we have to take into account the $\eta$-dependence
of $F$.

This is done by the basic solutions to the eigenvalue equation,
which is given by,
\be137 
F(\rho, \eta) = \Big[ \, { {\rm Im}\, \tau \over |m\tau +n|^2 }
\, \Big]^s = \Big[ { \rho \over m^2 \rho^2 + (m \eta +n) ^2 } \Big]^s
\quad , \qquad s = {3 \over 2} \, , \ -{1 \over 2}  \ . \\ 
\ee
For $m=0$ both the functions obviously reduce to the previous case 
and they are invariant under $SL(2, Z)$ transformations as given in eq.\
(\ref{521}) where $(m,n)$ transform as
\[
\Big( \ba{c} m \\ n \ea \Big) \rightarrow \Big(\ba{cc} a & c \\ b &d\ea
\Big) \Big( \ba{c} m \\ n \ea \Big) \ .
\]
Note, the equations are in fact $SL(2,R)$ invariant, but only for
$SL(2,Z)$ one can make a relation to an instanton sum. We see that the
$SL(2)$ transformations map the different solutions into each other
and summing over all co-prime integers $(m,n)$ gives the modular
invariant Eisenstein series $E_s$, which has been interpreted as a
summation over all (anti) instantons in \cite{925,922}. It is this
function which appears in the 4-d Einstein-Hilbert in the string frame
\cite{921}. Using the formula: $\Lambda(s) E_s = \Lambda(1-s) E_{1-s}$
with $\Lambda(s) = \pi^{-s} \Gamma(s) \zeta(2 s)$ \cite{270}, both
summations with $s={3 \over 2}$ or $s= -{1 \over 2}$ are equivalent.
The classical limit is obviously related to the value $s={3 \over 2}$,
which has as asymptotic expansion for large $\rho$
\be138
\ba{rcl}
4\pi E_{3/2}(\rho, \eta) & = & 2 \zeta (3)\rho^{3/2} +
{
2\pi^2\over 3} \rho^{-1/2} + 4\pi^{3/2} \sum_{m, n \geq 1}
{\Big ( {m \over n^3} \Big )}^{1/2} \nonumber \\[5mm]
&& \Big [ e^{2\pi imn (\eta +
i\rho)} +
e^{-2\pi imn(\eta - i\rho)} \Big ]
\Big [ 1 +
 \sum_{k=1}^{\infty} {\Gamma (k - 1/2)\over
\Gamma (-k -1/2)}{1\over (4\pi mn\rho)^k} \Big ]
\ea
\ee
that makes the tree level, one-loop and the non-perturbative
instanton contributions manifest.

We use the summation considered in \cite{924,915} yielding a
multi-pole solution of the self-dual Einstein metric by taking 
$s =-{1 \over 2}$ yielding
\be341
F = \sum_{(m,n)=1} {|m \tau  + n| \over \sqrt{{\rm Im} \tau }}
=\sum_{(m,n)=1} {\sqrt{m^2 \rho^2 + (m \eta +n)^2}
\over {\sqrt \rho}} 
\ee
where $(m,n)$ denotes the greatest common divisor.  The 3-pole
solution is particularly interesting and the corresponding family of
self-dual Einstein metrics have appeared in various contexts.  For
calculations below we consider this general 3-pole solution expressed
in the form,
\be342
F(\rho, \eta) = {a\over {\sqrt\rho}} + {b + c/q \over 2}
{\sqrt{\rho^2 + (\eta + 1)^2}\over {\sqrt\rho}} +
{b - c/q \over 2}{\sqrt{\rho^2 + (\eta - 1)^2}\over {\sqrt\rho}} \\
\ee
where, $a, b, c$ are some constants and $q^2 = \pm 1$ (in the
following we set $q=1$).  We shall consider such instanton corrections
and show that the gauged supergravity theory can admit de Sitter
vacua.  By making this truncation, we seem to have lost the classical
limit, i.e.\ the leading part in the expansion (\ref{138}) (but
recall, the series in $s=-{1 \over 2}$ and $s={3 \over 2}$ are
equivalent).  This can be justified, because the flow or the critical
points are on the line $\rho=0$, i.e.\ deep inside the quantum
region. But we have to keep in mind that this parameterization
also includes the homogeneous quaternionic space as special cases, 
namely
the Bergman metric, which also gives a parameterization of the coset
${SU(2,1) \over U(2)}$, is obtained for $a=c=1, \ b=0$ and the
hyperbolic space ${SO(4,1) \over SO(4)}$ corresponds to $b=c<0$
\cite{924}.


\section{Gauging the isometries, superpotential and potential}


We are primarily interested in de Sitter vacua appearing in gauged
supergravity, which may be unstable (saddle points), but the
unstable mode should flow into a supersymmetric vacuum.  Note, a
singular flow is not necessarily related to pole in the potential,
also simple run-away potentials yield a singular supergravity solution.
Therefore, we are looking for a potential that has two supersymmetric
local minima which are connected by a saddle point. 

The gauging of the coset space ${SU(2,1) \over U(2)}$  has
been considered in ref.\cite{913} and it has been shown
that as long as no vector multiplets are included the flow 
will always
end in a singularity. In the generic case, one obtains a run-away
behaviour, but there are also AdS as well as flat space vacua 
for this coset.

Let us first see whether the 1-loop correction can give additional
supersymmetric vacua, where the function $F$ is given by, 
\be141
F(\rho) = a {\rho}^{3/2} + {b \over {\sqrt{\rho}}}
\ee
with $a$ and $b$ as some constants and $b=0$ corresponds to the
classical case. Next, considering a general Killing vector like,
\be142  
k = c_1 \partial_{\phi} + c_2 \partial_{\psi} 
\ee
one finds that there is again only one supersymmetric fixed point at
$\rho^2 = {b\over a}$ and $\eta = -{c_2\over c_1}$. So in this case,
any non-supersymmetric de Sitter vacuum is of run-away type or runs
into a singularity.

However, considering the instanton corrections as discussed 
in the last section, one finds that one can have a rich vacuum 
structure and also good de Sitter vacua which is not the case 
for the classical and 1-loop corrected UH moduli space.   
For simplicity, we restrict ourselves to the 3-pole solution 
involving both $\rho$ and $\eta$ in a nontrivial way.  
  
Let us now consider gauging a linear combination of the 
two Abelian Killing vectors, namely,
\be140
k = \beta_1 \partial_\phi + \beta_2 \partial_\psi
\ee
where, $\beta_1$ and $\beta_2$ are two parameters. The norm 
of the Killing vector 
\be150
|k|^2 =  {1 \over F^2 \det Q} (\beta_1 \ \beta_2) \, N^t Q^2 N \,
\left(\ba{c} \beta_1 \\  \beta_2 \ea \right) 
\ee
has to vanish at fixed points which gives as  condition for
supersymmetric vacua
\be152
\det\Big| {N^t Q^2 N \over F^2 \det Q} \Big| = {1 \over F^4} = 0 \ .
\ee
For our choice of $F(\rho,\eta)$ in (\ref{341}), this corresponds to
$\rho=0$. But in fact both eigenvalues have to vanish which is the
case if: $|m \tau +n|=0$, i.e.\ at any given center, see also
\cite{924,915}. For this simple 3-pole solution we have exactly two
fixed points at
\be392
\rho = 0 \quad  , \qquad \eta = \pm 1 \ .
\ee
Note, these points are not regular in our coordinate system, but below
we will choose a regular coordinate system.

The SU(2) connections for the above CP metric are given by 
\cite{915}
\renewcommand{\arraystretch}{1.4}
\bea160
A^1 &=&- {\partial_\eta F \over F} \,  {d\rho}  + {1 \over \rho F}
\Big( {1 \over 2} F  + \rho \partial_\rho F \Big)\, {d \eta} \ , \\ 
A^2 &=& - {\sqrt{\rho} \over F} \, d\phi \ , \\
A^3 &=& {1 \over \sqrt{\rho} F} \, (d \psi + \eta d\phi)
\eea
and, with the above Killing vector, the Killing prepotentials
are obtained as,
\be170
P^1 = 0 \quad , \qquad 
P^2 = -\beta_1 \, {\sqrt{\rho} \over F} \quad , \qquad
P^3 = {1 \over \sqrt{\rho} F} \, (\beta_2 + \eta \beta_1) \ .
\ee
The superpotential therefore reads, 
\be180
W \equiv \sqrt{P^iP^i}=  {\sqrt{\beta_1^2 \rho^2 + 
(\beta_2 + \eta \beta_1 )^2} \over \sqrt{\rho} F} \ .
\ee
The different supersymmetric vacua have been discussed already in
\cite{915}, see also \cite{280} for a different parameterization of
the 3-pole solution. In this paper we are interested in de Sitter vacua
emerging between two supersymmetric vacua and which is possible if
both supersymmetric vacua correspond to local minima of the
supergravity potential. To make sure that the value of the potential
at the saddle is positive, we choose one of the supersymmetric vacua
as flat space. Note, a supersymmetric flat space vacuum has a positive
definite mass matrix (at least as long the moduli metric does not
become degenerate at this point) and hence, choosing one supersymmetric
vacuum as flat space, there has to be a deSitter saddle point
in-between. If we take $\beta_1 = 1$ and $\beta_2 = -1$ for
the Killing vector, one obtains the superpotential as 
\be200
W = {\sqrt{\rho^2 + (\eta -1)^2}\over \sqrt{\rho} F} \ .
\ee
which then vanishes at the critical point $\rho=0\ , \ \eta = 1$.

Let us now introduce new coordinates
\be210
\rho = \sinh r \, \cos\theta \quad ,\qquad \eta= \cosh r \,
\sin\theta \ .
\ee
giving
\bea111
{1 \over 4\, \rho} \Big[F - 4 \rho^2 (F_\rho^2 + F_\eta^2)\Big]
&=& {b^2 -c^2 + a (b \cosh r -c \sin\theta) \over \cosh^2 r - \sin^2\theta}
\nonumber
\\[4mm]
\sqrt{\rho} F &=& a + b \cosh r + c \sin\theta  \ . \nonumber
\eea
For the flow equations, only the ($\rho, \eta$) part of the metric
matters and in the new coordinates, this part becomes
\be212
{c^2 -b^2 - a (b \cosh r -c \sin\theta) \over (a + b \cosh r
+ c \sin\theta)^2} \Big( dr^2 + d \theta^2 \Big) \ .
\ee
We have changed here the sign of the metric as we are interested
in negatively curved anti-selfdual Einstein space, see discussion after
eq.\ (\ref{110}). In these coordinates, the superpotential becomes 
\be220
W = {\cosh r  - \sin\theta \over a + b \cosh r  + c \sin\theta}
\ee
yielding fixed points at
\be230
r =0  \quad , \qquad \sin\theta= \pm 1 \ .
\ee
So, we see that $\theta$ can run between a flat space vacuum ($r=0,
\theta = \pi/2$) where $W = dW = 0$ and an AdS vacuum ($r=0, \theta =
-\pi/2$) where $W \neq 0, \ dW = 0$.  Note, with this change of
coordinates the metric at the fixed points is well defined, but there
is still a coordinate singularity at $\sqrt{\rho} F = a + b \cosh r +
c \sin\theta = 0$, which represents a pole in the superpotential and
hence the supergravity potential $V$ is not bounded from below. But 
by fixing the parameter appropriately this pole will be separated by a
potential barrier. On the other hand, the moduli space has a
singularity at $\det Q= 0$, which in the new coordinates corresponds
to the point where $c^2 - b^2 - a (b \cosh r -c \sin\theta)=0$\, .
Recall that the sign of ($\det Q$) fixes the sign of the curvature of
the 4-dimensional space and hence we have to restrict ourselves to
regions where $\det Q < 0$ and drop the overall sign in the metric
[this is what we have done already in (\ref{212})].

In order to investigate the possibility of  de Sitter vacua, we
have to discuss the supergravity potential, which is given by
\be240
V  = 2 \Big[ g^{rr} (\partial_r W)^2 + g^{\theta\theta}
(\partial_\theta W)^2 -  {3 \over 2} W^2 \Big] \ .
\ee
At the two BPS extrema ($r =0 , \ \theta = \pm {\pi \over 2}$), the
eigenvalues of the Hessians of $W$ and $V$ are given by,
\be250
\ba{lcl}
ddW\Big|_{\theta={\pi \over 2}} &=& \  \Big[ \; {1 \over a+b+c}\ , \ 
 {1 \over a+b+c}\ \Big] \\
ddV\Big|_{\theta={\pi \over 2}} &=&  4 \Big[ \ 
 {  1 \over (c-b)(a+b+c)} \ , \  
 { 1 \over (c-b)(a+b+c)}\ \Big] \\
ddW\Big|_{\theta=-{\pi \over 2}} &=& \  \Big[\ {a-b-c \over (a+b-c)^2}
\ , \ - {a+b+c \over (a+b-c)^2} \Big] \\ 
ddV\Big|_{\theta=-{\pi \over 2}} &=& 4 \Big[\ 
{(2[b+c] +a)(b+c-a)  \over (b+c)(a+b-c)^3}
\ , \  {(2[b+c] -a)(a+b+c)  \over (b+c)(a+b-c)^3}
\Big]
\ea
\ee 
with $ W\Big|_{\theta= - {\pi \over 2 }} = {2 \over a +b -c}$.
As imposed by supersymmetry, the masses of the scalars are the same in
the flat space vacuum at $(r=0 , \ \theta = {\pi \over 2})$.

As one might have expected for this inhomogeneous space, the metric
by itself is not positive definite, there are different regions in
parameter space where the metric has the correct signature -- one of
them will contain the singularity. We now want to choose the parameter
in a way, that this region containing both supersymmetric fixed points,
is regular and has the correct signature. Therefore, we have to impose
the following relations
\be726
c^2 - b^2 - a (b \pm c) = -(b\pm c)(a+ b \mp c ) > 0 
\  , \quad
a + b \mp c < 0 \    
\ee
yielding $b \pm c > 0$ or
\be611
c^2 - b^2 - a (b \pm c) = -(b\pm c)(a + b \mp c) > 0 
\  , \quad
a + b \mp c > 0 \   . 
\ee
giving $b \pm c < 0$. The first condition ensures that the metric
(\ref{212}) at both BPS fixed points is positive definite and the
second condition implies that there is no coordinate singularity
in-between, i.e.\ $\sqrt{\rho} F$ does not change its sign. It is
straightforward to verify that these conditions also ensure that the
flat space is a local minimum.  To simplify the notation further let
us set $b=\pm 1$ which gives the new relations
\be662
\ba{rcl}
& b=1 \quad , \qquad a < -2 \quad , \qquad |c|>1 & ,\\
& b=-1 \quad , \qquad a >  2 \quad , \qquad |c| <1 &  .
\ea
\ee
In the figure we have plotted an example for the potential, where the
BPS vacua are local minima and in-between is a deSitter saddle point
(it is of course $2 \pi$-periodic in $\theta$).  In the flat space
vacuum the mass matrix is always positive definite, but in the anti
deSitter vacuum $\partial_r \partial_r V$ is positive only if $a > 2|
b+c|$ (note $b+c < 0 $ in this example), but even if this does not hold,
this vacuum is stable due to the Breitenlohner-Freedman bound
saturated by any BPS fixed point.  Moreover, we see that the critical
line $r=0$ is a local minimum, but for large $r$ we always run into a
pole of the superpotential, where
\[
a + b \cosh r + c \sin\theta = 0 
\]
(this point in moduli space is smooth, but it represents the endpoint
of a given coordinate region). In the supergravity potential $V$, a
pole in $W$ can result in a positive or negative pole depending on
whether the first term grows faster than the second. Unfortunately, in
our case it is negative pole, which can be seen as follows.  In
addition to the critical line: $r=0$, also the lines: $\theta = \pm
{\pi \over 2}$ are critical (ie.\ $\partial_\theta V\big|_{\theta =
\pm {\pi \over 2}} =0$ for all values of $r$). Let us now investigate
the line $\theta = {\pi \over 2}$, where the potential around the flat
space vacuum  is positive definite and vanishes at $r =0$. It is a
straightforward exercise to see that the potential along this line as
function of $\cosh r$ has two further zeros at
\[
\ba{rcl}
\cosh r_\pm &=& -{1 \over 6ab} \Big[5 b^2-c^2+ 2a^2 +a (c+ b) + 4 c b \hfill \\
  && \qquad \pm
\sqrt{(25 b^2+4 a^2+c^2- 28 a b-10 c b-4 a c) (c+a+b)^2)} \Big] \ .
\ea
\]
The rhs is exactly equal to one at: $a+b+c=0$, but for all other
values discussed in eqs.\ (\ref{726}) or (\ref{611}), only one of the
solutions obeys: $\cosh r_\pm >1$. This implies, that the potential
$V|_{\theta = {\pi \over 2}}$ has, apart from the point at $r=0$,
always one additional zero. Hence, there is a {\em negative} pole in
$V$ and there is another  de Sitter vacuum, where $V$ reaches its
maximum. The same analysis can be repeated for the critical line at
$\theta = -{\pi \over 2}$. If the BPS vacuum at $r=0$ is a local
minimum we get another de Sitter saddle point at the edge of the
negative pole.  Note, the existence of these two additional de Sitter
extrema is a simple consequence of the positivity of the Hessian of
$V$ at each BPS extremum and the appearance of the negative pole in
the potential for large values of $r$.

\begin{figure} 
\begin{center}
\includegraphics[angle=-90,width=100mm]{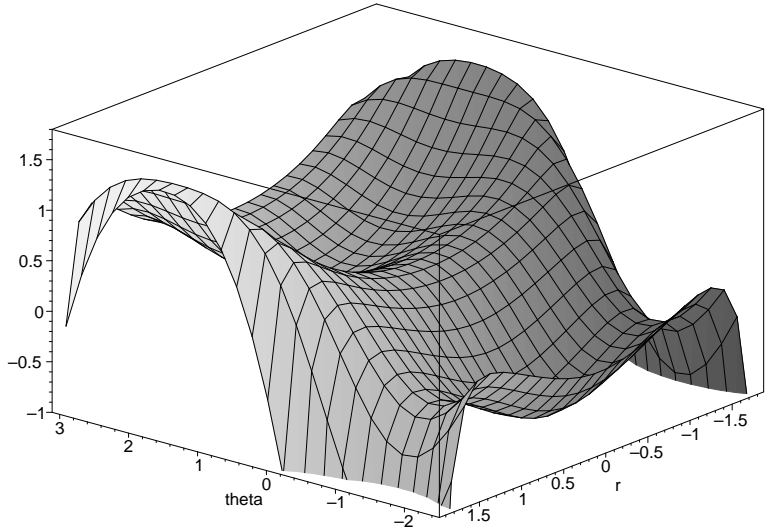}
\end{center}
\caption{For this plot we have chosen $a=5$ ,  $b =-1$ , $c=0.1$ .
It shows one AdS minimum at $(r, \theta) = (0, -{\pi \over 2})$,
a flat space vacuum at  $(r, \theta) = (0, {\pi \over 2})$ and
in-between there is a deSitter saddle point.}
\label{figure1}
\end{figure}

Recall that the appearance of {\em two} BPS extrema was a consequence of the
3-pole ansatz in (\ref{342}) and each additional pole results into
another supersymmetric extremum, but of course not all of them are
connected, i.e.\ they should  be separated by poles of the superpotential.
In fact as we know from the RG-flow discussion, smoothly connected BPS
flows are only possible between the so-called UV (ultraviolet) and IR 
(infrared) or IR and IR
fixed points, see e.g.\ \cite{913}. All BPS fixed points are on the
line $r=0$ and are separated in the $\eta$-direction and therefore the
$\eta$-dependence was crucial in our setup -- the 1-loop correction to
the classical result is not sufficient.
 
Let us end this section, with a discussion for the case when both BPS
vacua are AdS, which is the case if we choose for the Killing vector:
$\beta_1 = 1$ and $\beta_2 = \lambda \neq \pm 1$. The superpotential
for such case in $r$ and $\theta$ coordinates is given by,
\be127
W = {\sqrt{-\cos^2 \theta + \cosh^2 r + 2 \cosh r\sin\theta
\lambda + \lambda^2} \over a + b\cosh r + c\sin \theta}
\ee
The BPS critical points are again at $r=0$ and $\cos\theta =0$
which gives
\be128
\ba{l}
W\Big |_{r=0, \theta={\pi \over 2}} = {\lambda + 1 \over a + b + c} \\
W\Big |_{r=0, \theta=-{\pi \over 2}} = {\lambda - 1 \over a + b - c} \\
\ea
\ee
and for $\lambda = \pm 1 $ we get back to our previous case of flat
space and AdS vacuum. Choosing $|\lambda| > 1$, one fixed point is of
UV and the other is of IR type and for $|\lambda|<0$ both are of IR type, 
which is the relevant setup for the Randall-Sundrum scenario \cite{906}.
Since the $r$ direction is stable ($\partial_r \partial_r V > 0$), one
can consider the case where the scalar $r$ is frozen to its fixed
point value. One then gets,
\be129
W = {\lambda + \sin\theta \over a + b + c\sin\theta}
\ee
The potential $V$ is then given by,
\be130
V = 2 \Big [ {\cos^2\theta \, (a + b - c\lambda)^2 \over 
[c^2 - b^2 - a(b - c\sin\theta)]
(a + b + c\sin\theta)^2} - {3 \over 2}{(\lambda + \sin\theta)^2 \over
(a + b + c\sin\theta)^2} \Big ]
\ee
%


\section{ Cosmological flow towards the flat space minimum}


Let us go back to the case where we have the flat space and AdS vacua
as the two supersymmetric extrema (for the Killing vector with the
choice $\beta_1 =1$ and $\beta_2 = -1$). As we saw before, the non
supersymmetric vacuum corresponding to the de Sitter saddle point had
an unstable direction along $\theta$. Since the solution is stable
along the $r$ direction, we can freeze the $r$ coordinate to its fixed
point value so that there is only one scalar field $\theta$. One can
now look for a time dependent solution departing from the de Sitter
extremum along the $\theta$ direction to reach the supersymmetric flat
space minimum. For this, we have to consider the scalar field $\theta$
having a time dependence and we make an ansatz for a time dependent
flat Robertson-Walker metric in four dimensions, which is given by,
\be523
ds^2 = - dt^2 + e^{2a(t)}\Big (dx^2 + dy^2 + dz^2 \Big ) \ .
\ee
where $a(t)$ is the scale factor.

If one now starts at the de Sitter saddle point and runs towards the
AdS minimum, then with the above metric, the system can never 
settle down in this
BPS vacuum, which is a consequence of the Einstein equations and has
been used in the fast roll inflation proposal \cite{190}, see also
\cite{930}. However, one has to keep in mind that this conclusion is
a consequence of the metric ansatz and allowing for a non-flat spatial
section and/or allowing for a dependence on the spatial 
coordinates,
the system may settle down in the AdS minimum. But the above metric
ansatz is phenomenologically most interesting and hence let us
stick to this ansatz.

The action involving the scalar field $\theta$ and the potential
$V(\theta)$ is given by,
\be524
S \sim \int dt d^3x {\sqrt{-g}}
\Big ( R - {1\over 2}g_{\theta\theta}g^{\mu\nu}\partial_\mu\theta
\partial_\nu\theta - V(\theta) \Big )
\ee
where, $\mu, \nu$ are 4-dimensional space-time indices. 
The equations of motion become
\be525
\ba{rcl}
3 {\dot a}^2  - {1\over 4}g_{\theta\theta}{\dot\theta}^2 -
{1\over 2}V(\theta)  &=&  0 \\
-2 \ddot a - 3 {\dot a}^2 - {1 \over 4}g_{\theta\theta}
{\dot\theta}^2 + {1 \over 2} V(\theta)  &=&  0 \\
g_{\theta\theta}(\ddot\theta + 3\dot a\dot\theta) +
\partial_{\theta} V(\theta)
+ {1\over 2} (\partial_{\theta}g_{\theta\theta}) {\dot\theta}^2 &=& 0
\ea
\ee
where the last equation is the scalar equation of motion.
Adding the first two equations, one gets,
\[
\ddot a = -{1\over 4}g_{\theta\theta}(\dot\theta)^2 \\
\]
which is also known as the cosmological c-theorem (i.e.\ $\dot a$ is a
monotonic function).  Then differentiating the first equation with
respect to time, one gets,
\be528
6\dot a \ddot a - {1\over 2} g_{\theta\theta} \dot\theta \ddot\theta
-{1\over 4} \partial_{\theta}g_{\theta\theta} (\dot\theta)^3 -
{1\over 2}\dot\theta \partial_{\theta} V = 0 \\
\ee
Inserting the value of $\partial_{\theta} V$ from the scalar field
equation, and using $\ddot a = -{1\over 4} g_{\theta\theta}
(\dot\theta)^2$ in the above equation, one finds that one of the
metric equation is redundant.  Hence one has to solve only the scalar
equation and the $g_{00}$ equation to obtain time dependent solutions
for $\theta$ and the scale factor.  Note that, these two equations do
not reduce to first order equations as has been the case for the BPS
domain wall solutions. Though we have not solved these equations
explicitly, qualitatively one finds that the solution asymptotes to
the flat space where $W=dW=0$ which means that the scalar field rolls
down from the de Sitter saddle point and stabilizes at the minimum.

Of course the appearance of the negative pole in the potential
questions the model from the phenomenological perspective. It would
have been much better, if one had found a parameterization so that $V$
was bounded from below. Recall, poles in the superpotential can also
yield positive poles in $V$ for the case that first term grows faster
than the second. Unfortunately, we were not able to find such a setup.
On the other hand, one can of course identify the Big Bang with this
pole and with the correctly tuned initial velocity, the scalars can
reach the de Sitter saddle point and finally settle down in the flat
space vacuum.


\section{Discussion}


In this paper, we have studied the gauging of the isometries of the
quantum corrected universal hypermultiplet moduli space, where the
corresponding quaternionic geometry has been given by the Calderbank -
Pedersen metric. We have analyzed the superpotential and the potential
to obtain the fixed points relevant in the context of holographic RG
flow. We have considered the simplest 3-pole solution which
essentially includes the non-perturbative instanton corrections and we
have explored the possibility of obtaining de Sitter vacua with such a
solution.  We have found a de Sitter saddle point that connects two
supersymmetric minima of the potential.  Moreover, the potential has
one de Sitter maximum as well as another de Sitter saddle point, which
are however at the edge of a negative pole in the potential.
  
The dS/CFT correspondence \cite{927} has led to the conjecture that
the Universe is an RG flow between two conformal fixed points of a
three dimensional Euclidean field theory where the time evolution in
the bulk has been interpreted as inverse RG flow in the dual CFT
\cite{928}. In ref.\cite{930}, the cosmological evolution of a scalar
field coupled to gravity has also been studied and a detailed analysis
of the extrema of the potential and the flow equations have been done.
Our case is different from these considerations as the flow considered
here does not really have a field theory interpretation.  Note that
the flow from the de Sitter saddle point to the AdS vacua will be
ruled out as the (time-dependent) scalars can not settle in the AdS
minimum and only a flow towards flat space minimum is allowed (at
least as long as one sticks to the spatially flat Robertson-Walker
metric).  We also should mention here that if we consider the case
where we have two AdS supersymmetric vacua, then with the above metric
ansatz, there should be only bubble solution where starting from the
dS vacuum, one runs to the another dS vacuum which will be more in
line with the dS/CFT correspondence.

Finally we would like to comment on the relevance of our
superpotential to the recently discussed new-old inflationary models
\cite{931} where the potential does not have to satisfy the slow roll
condition. There, it has been possible to get inflation when the
system is locked at a saddle point by sufficient oscillation along the
stable direction and eventually the scalar rolls down to the true
minimum of the potential.  It might be interesting to see whether the
de Sitter saddle point can be used for a locked inflation before the
system settles down in the flat space minimum.  The scalar potential
depending on two scalar fields $r$ and $\theta$, when expanded up to
second order and choosing for example, the constant parameters $a, b,
c$ satisfying the condition to avoid the curvature singularity, one
can put it into a form as in the new old inflationary scenario.  Our
potential has however no false vacuum and hence, from the cosmological
point of view, the evolution can start with the singularity
corresponding to the negative pole in the potential. But then in order
to reach the saddle point and initiate a locked inflation, scalar fields
have to have fine-tuned initial velocities. A more detailed
investigation along these lines will be very interesting.

{\bf Acknowledgments}

We would like to thank G. Dall'Agata, S. Gukov, S. Minwalla, 
L. Motl, A. Sen, A. Strominger for interesting discussions.
S.M. would like to thank the High Energy Physics group at 
Harvard University and especially Shiraz Minwalla for the 
warm hospitality where a part of this work has been done. 
S.M. also acknowledges I.O.P. Bhubaneswar for extending computer
facilities. The work of K.B. is supported by a Heisenberg grant
of the DFG.



\begin{thebibliography}{10}

\bibitem{900}
J.~Maldacena, ``The large N limit of superconformal field theories
and supergravity'', {\em Adv. Theor. Math. Phys.} {\bf 2} (1998) 231 
[hep-th/9711200].
 
\bibitem{801}
M.~Cveti{\v c} And H.~H.~Soleng, ``Supergravity domain walls'', {\em
Phys. Rep.} {\bf 282} (1997) [hep-th/9604090].

\bibitem{905}
D.~Z.~Freedman, S.~S.~Gubser, K.~Pilch and N.~P.~Warner,
``Renormalization group flows from holography sypersymmetry and
a c-theorem'', {\em Adv. Theor. Math. Phys.} {\bf 3} (1999) 363
[hep-th/9904017];
K.~Behrndt and M.~Cveti{\v c}, ``Supersymmetric domain-wall world
from D=5 simple gauged supergravity'', {\em Phys. Lett.} {\bf B}
(2000) 253 [hep-th/9909058];
R.~Kallosh, A.~Linde and M.~Shmakova, ``Supersymmetric multiple
basin attractors'', {\em JHEP} {\bf 9911} (1999) 010 [hep-th/9910021].
R.~Kallosh and A.~Linde, ``Supersymmetry and brane world'', {\em JHEP}
{\bf 0002} (2000) 005 [hep-th/0001071];
K.~Behrndt and M.~Cveti{\v c}, ``Anti de Sitter vacua of gauged 
supergravities
with 8 supercharges'', {\em Phys. Rev.} {\bf D61} (2000) 101901
[hep-th/0001159].

\bibitem{906}
L.~Randall and S.~Sundrum, ``An alternative to compactification'',
{\em Phys. Rev. Lett.} {\bf 83} (1999) 4690 [hep-th/9906064].

\bibitem{907}
M.~Gunaydin, G.~Sierra and P.~K.~Townsend, ``Gauging the D=5 Maxwell-
Einstein Supergravity theories:More on Jordan algebra'',
{\em Nucl. Phys.} {\bf B253} (1985) 573;
A.~Ceresole and G.~Dall'Agata, ``General mather coupled N=2,
D=5 gauged supergravity'', {\em Nucl. Phys.} {\bf B585} (2000)
143 [hep-th/0004111];
M.~Gunaydin and M.~Zagermann, ``The vacua of 5d, N=2 gauged
Yang-Mills/Einstein/tensor supergravity: Abelian case'',
{\em Phys. Rev.} {\bf D62} (2000) 044028 [hep-th/0002228];
L.~Andrianopoli, M.~Bertolini, A.~Ceresole, R.~D'Auria,
S.~Ferrara, P.~Fre and T.Magri, ``N=2 supergravity and N=2
super Yang-Mills theory on general scalar manifolds: Symplectic
covariance, gaugings and the momentum map'', {\em J. Geom,
Phys.} {bf 23} (1997) 111 [hep-th/9605032].

\bibitem{911}
A.~Lukas, B.~A.~Ovrut, K.~S.~Stelle and D.~Waldram, ``Heterotic
M-theory in five dimensions'', {\em Nucl. Phys.}{\bf B552} (1999)
246 [hep-th/9806051];

\bibitem{913}
K.~Behrndt, C.~Herrmann, J.~Louis and S.~Thomas,
``Domain walls in five dimensional supergravity with nontrivial
hypermultiplets'', {\em JHEP} {\bf 01} (2001) 011 [hep-th/0008112];
K.~Behrndt and M.~Cveti{\v c}, ``Gauging of N=2 supergravity hypermultiplet
and novel renormalization group flows'', {\em Nucl. Phys.} {\bf 609}
(2001) 183 [hep-th/0101007];
A.~Ceresole, G.~Dall'Agata, R.~Kallosh and A.~Van~Proyen,
``Hypermultiplets, Domain walls and Supersymmetric Attractors'',
{\em Phys. Rev.} {\bf D64} (2001) 104006 [hep-th/0104056].


\bibitem{915}
L.~Anguelova and C.~I.~Lazaroiu, ``Domain walls of N=2 supergravity
in five dimensions from hypermultiplet moduli spaces'',
{\em JHEP} {\bf 0209} (2002) 053 [hep-th/0208154].

\bibitem{916}
K.~Behrndt and G.~Dall'Agata, ``Vacua of N=2 gauged supergravity
derived from non-homogeneous quaternionic spaces'', {\em Nucl.
Phys.} {\bf B627} (2002) 357 [hep-th/0112136].

\bibitem{917}
R.~D'Auria and S.~Ferrara, ``On fermion masses, gradient flows and
potential in supersymmetric theories'', {\em JHEP} {\bf 05}
(2001) 034 [hep-th/0103153].

\bibitem{260}
E.~Bergshoeff, M.~de Roo and E.~Eyras,
``Gauged supergravity from dimensional reduction,''
{\em Phys.\ Lett.} {\bf B 413}, 70 (1997)
[hep-th/9707130];
K.~Behrndt, E.~Bergshoeff, D.~Roest and P.~Sundell,
``Massive dualities in six dimensions,''
{\em Class.\ Quant.\ Grav.} {\bf 19}, 2171 (2002)
[hep-th/0112071].


\bibitem{180}
M.~de Roo, D.~B.~Westra and S.~Panda,
``De Sitter solutions in N = 4 matter coupled supergravity,''
{\em JHEP} {\bf 0302}, 003 (2003)
[hep-th/0212216];
M.~de Roo, D.~B.~Westra, S.~Panda and M.~Trigiante,
``Potential and mass-matrix in gauged N = 4 supergravity,''
{\em JHEP} {\bf 0311}, 022 (2003)
[hep-th/0310187].



\bibitem{160}
N. Ohta, ``Intersection rules for S-branes", {\em Phys. Lett.} {\bf B558}
(2003) 213 [hep-th/0301095]; P.~K.~Townsend and M.~N.~R.~Wohlfarth,
``Accelerating cosmologies from compactification'',
{\em Phys.\ Rev.\ Lett.} {\bf 91} (2003) 061302
[hep-th/0303097];
N.Ohta, ``Accelerating cosmologies from S-branes'', 
{\em Phys. \ Rev.\ Lett.} {\bf 91} (2003) 061303-1 [hep-th/0303238];
``A study of accelerating cosmologies 
from Superstring/M Theories'', {\em Prog. Theor. Phys.} {\bf 110}
(2003) 269 [hep-th/0304172]. 


\bibitem{210}
C.~M.~Hull,
``De Sitter space in supergravity and M theory,''
{\em JHEP} {\bf 0111}, 012 (2001)
[hep-th/0109213];
G.~W.~Gibbons and C.~M.~Hull,
``de Sitter space from warped supergravity solutions,''
[hep-th/0111072]. 


\bibitem{150}
R.~Kallosh, A.~D.~Linde, S.~Prokushkin and M.~Shmakova,
``Gauged supergravities, de Sitter space and cosmology,''
{\em Phys.\ Rev.} D {\bf 65} (2002) 105016
[hep-th/0110089].


\bibitem{100}
P.~Fre, M.~Trigiante and A.~Van Proeyen,
``Stable de Sitter vacua from N = 2 supergravity,''
{\em Class.\ Quant.\ Grav.}  {\bf 19}, 4167 (2002)
[hep-th/0205119].

\bibitem{211}
N.~S.~Deger, ``Renormalization Group flows from D=3, N=2 matter 
coupled gauged supergravities'', {\em JHEP} {\bf 0211}
(2002) 025 [hep-th/0209188]. 
\bibitem{240}
R.~Blumenhagen, B.~K\"ors, D.~L\"ust and T.~Ott,
``Hybrid inflation in intersecting brane worlds,''
{\em Nucl.\ Phys.} {\bf B 641}, 235 (2002)
[hep-th/0202124].

\bibitem{120}
S.~Kachru, R.~Kallosh, A.~Linde, J.~Maldacena, L.~McAllister and 
S.~P.~Trivedi,
``Towards inflation in string theory,''
{\em JHEP} {\bf 0310} (2003) 013
[hep-th/0308055];
S.~Kachru, R.~Kallosh, A.~Linde and S.~P.~Trivedi,
``De Sitter vacua in string theory,''
{\em Phys.\ Rev.} {\bf D} {\bf 68}, 046005 (2003)
[hep-th/0301240].


\bibitem{110}
C.~P.~Burgess, R.~Kallosh and F.~Quevedo,
``de Sitter string vacua from supersymmetric D-terms,''
{\em JHEP} {\bf 0310} (2003) 056
[hep-th/0309187].


\bibitem{333}
G.~Dall'Agata, unpublished notes, 2003.


\bibitem{190}
A.~Linde,
``Fast-roll inflation,''
{\em JHEP} {\bf 0111}, 052 (2001)
[hep-th/0110195].


\bibitem{931}
G.~Dvali and S.~Kachru, ``New old inflation'', [hep-th/0309095];
``Large scale power and running spectral index in new
old inflation'', [hep-ph/0310244];
G.~Dvali and S.~Kachru,
``Large scale power and running spectral index in new old inflation,''
[hep-ph/0310244].


\bibitem{140}
M.~Axenides and K.~Dimopoulos,
``Inflation without flat directions,''
[hep-ph/0310194].


\bibitem{918}
A.~Strominger,``Loop corrections to the universal hypermultiplet'',
{\em Phys. Lett.} {\bf B421} (1998) 139 [hep-th/9706195].

\bibitem{919}
M.~B.~Green, M.~Gutperle and P.~Vanhove, ``One loop in eleven
dimension'', {\em Phys. Lett.} {\bf B409} (1997) 177 [hep-th/9706175];
I.~Antoniadis, S.~Ferrara, R.~Minasian and K.~S.~Narain,
``$R^4$ couplings in M-theory and type-II theories in Calabi-Yau spaces'',
{\em Nucl. Phys.} {\bf B507} (1997) 571 [hep-th/9707013].

\bibitem{921}
I.~Antoniadis,R.~Minasian, S.~Theisen and P.~Vanhove, ``String
loop corrections to the universal hypermultiplet'', {\em Class. 
Quant. Grav.} {\bf 20} (2003) 5079 [hep-th/0307268].

\bibitem{922}
S.~Ketov, ``Non perturbative low energy effective action of the
universal hypermultiplet'', [hep-th/0301074];
S.~Ketov, ``Summing up D-instantons in N=2 supergravity'', {\em Nucl.
Phys.} {\bf B649} (2003) 365 [hep-th/0209003].


\bibitem{923}
U.~Theis and S.~Vandoren, ``Instantons in the double-tensor multiplet'',
{\em JHEP} {\bf 0209} (2002) 059 [hep-th/0208145];
M.~Davidse, M.~de Vroome, U.~Theis and S.~Vandoren,
``Instanton solutions for the universal hypermultiplet,''
[hep-th/0309220].

\bibitem{924}
D.~M.~J.~Calderbank and H.~Pedersen, ``Self-dual Einstein metrics with
torus symmetry'', {\em J. Diff. Geom.} {\bf 60} (2002) 485
[math.DG/0105263].


\bibitem{230}
J.~Louis,
``Aspects of spontaneous N = 2 $\to$ N = 1 breaking in supergravity,''
[hep-th/0203138].


\bibitem{925}
M.~B.~Green and M.~Gutperle, ``Effects of D-instantons'', {\em
Nucl. Phys.} {\bf B498} (1997) 195 [hep-th/9701093];
M.~B.~Green and S.~Sethi,
``Supersymmetry constraints on type IIB supergravity,''
{\em Phys.\ Rev.} {\bf D 59}, 046006 (1999)
[hep-th/9808061].


\bibitem{270}
A. Terras,
``Harmonic analysis on symmetric spaces and applications I''
{\em Springer, Berlin -- New York}, 1985.


\bibitem{280}
K.~Behrndt, G.~Dall'Agata, D.~L\"ust and S.~Mahapatra,
``Intersecting 6-branes from new 7-manifolds with G(2) holonomy,''
{\em JHEP} {\bf 0208} (2002) 027
[hep-th/0207117].

\bibitem{930}
R.~Argurio, ``Comments on cosmological RG flows'', {\em JHEP}
{\bf 0212} (2002) 057 [hep-th/0202183].



\bibitem{927}
A.~Strominger, ``The dS/CFT correspondence'', {\em JHEP} {\bf 0110}
(2001) 034 [hep-th/0106113]; M.~Spradlin, A.~Strominger, A.~Volovich,
``Les Houches lectures on de Sitter space'' [hep-th/0110007].

\bibitem{928}
A.~Strominger, ``Inflation and the dS/CFT correspondence'',
{\em JHEP} {\bf 0111} (2001) 049 [hep-th/0110087].




\end{thebibliography}

\providecommand{\href}[2]{#2}\begingroup\raggedright\endgroup


\end{document}